\documentclass[useAMS,usenatbib]{mn2e}

\usepackage{Times}
\usepackage{graphicx}

\title[A progenitor for SN 2007bi]{A progenitor for the extremely luminous Type Ic supernova 2007bi}
\author[T. Yoshida and H. Umeda]{T. Yoshida\thanks{E-mail:
tyoshida@astron.s.u-tokyo.ac.jp} and H. Umeda\\
Department of Astronomy, Graduate School of Science, University of Tokyo, Tokyo 113-0033, Japan}
\begin{document}

\date{Accepted 2010 December 29. Received 2010 December 29; in original form 2010 November 11}

\pagerange{\pageref{firstpage}--\pageref{lastpage}} \pubyear{2010}

\maketitle

\label{firstpage}

\begin{abstract}
SN 2007bi is an extremely luminous Type Ic supernova.
This supernova is thought to be evolved from a very massive star, and
two possibilities have been proposed for the explosion mechanism.
One possibility is a pair-instability supernova with 
an $M_{{\rm CO}} \sim 100$ M$_\odot$ CO core progenitor.
Another possibility is a core-collapse supernova with
$ M_{{\rm CO}} \sim 40$ M$_\odot$.
We investigate the evolution of very massive stars with main-sequence mass
$M_{{\rm MS}} = 100 - 500$ M$_\odot$ and $Z_0 = 0.004$, which is 
in the metallicity range of the host galaxy of SN 2007bi, to constrain 
the progenitor of SN 2007bi.
The supernova type relating to the surface He abundance is also discussed.
The main-sequence mass of the progenitor exploding as a pair-instability
supernova could be $M_{{\rm MS}} \sim 515 - 575$ M$_\odot$.
The minimum main-sequence mass could be 310 M$_\odot$ when uncertainties
in the mass-loss rate are considered.
A star with $M_{{\rm MS}} \sim 110 - 280$ M$_\odot$ evolves to a CO star,
appropriate for the core-collapse supernova of SN 2007bi.
Arguments based on the probability of pair-instability and 
core-collapse supernovae favour the hypothesis
that SN 2007bi originated from a core-collapse supernova event.

\end{abstract}


\section{Introduction}

SN 2007bi was found as an extremely luminous Type Ic supernova (SN Ic).
Spectral analyses deduced the production of $3.6 - 7.4$ M$_\odot$ of
radioactive $^{56}$Ni \citep{Gal-Yam09}.
The metallicity of the host galaxy of SN 2007bi, which is a subluminous
dwarf galaxy, has been observed to be $12+\log({\rm O/H}) = 8.15 \pm 0.15$, 
corresponding to $Z=0.2-0.4$ ${\rm Z}_\odot$ \citep{Young10}.
This metallicity is similar to those of low-metallicity galaxies undergoing
$\gamma$-ray bursts (GRBs) associated with SNe.

This SN was proposed as a pair-instability (PI) SN from observations of 
the light curve and spectral analyses.
The final mass was estimated to be $\sim 100$ M$_\odot$.
On the other hand, the light curve was also fitted by the energetic 
core-collapse (CC) explosion model \citep{Moriya10}.
The estimated progenitor was a $43-$M$_\odot$ CO core.
40 M$_\odot$ of ejecta, containing 6.1 M$_\odot$ of 
$^{56}$Ni, were thought to be ejected.
Although there is a difference in the rise period of the light curves
of these models, there were no observations during the period.
Hence, the explosion mechanism of SN 2007bi has not yet been clarified.
Observational features such as the total $^{56}$Ni mass and the SN type 
will provide constraints for the progenitor.

The evolution of very massive stars connects presupernova structures
and stars at the main-sequence (MS) stage.
The final stellar mass, the mass of the CO core and the surface composition
depend on MS mass and metallicity through burning processes and 
mass loss.
Calculations of the evolution of very massive stars will provide the 
relationship of these features to MS mass.

The relationship between the amount of $^{56}$Ni produced in a SN and 
the CO core mass has been evaluated for PI and CC SN models.
The mass range of the CO cores for PI SN progenitors has been evaluated as
$M_{{\rm CO}} \sim 60 - 130$ M$_\odot$ \citep{Heger02,Umeda02}.
Smaller progenitors for PI SNe were also suggested by \citet{Waldman08}.
The amount of $^{56}$Ni produced in a PI SN increases with MS mass 
\citep{Heger02,Umeda02}.
The amount of $^{56}$Ni ejected from a CC SN is related to the CO core 
mass and the explosion energy \citep{Umeda08}.
A SN with an $M_{{\rm CO}} \ga 35$ M$_\odot$ progenitor can produce
more than 3 M$_\odot$ of $^{56}$Ni when the explosion energy is
$E_{{\rm kin}} = 3 \times 10^{52}$ erg.
These relations connect the amount of $^{56}$Ni deduced from
the observations of SN 2007bi with the theoretical MS mass of a very massive 
star.

In this Letter, we evaluate the mass range of very massive stars with
initial metallicity of $Z_0 = 0.004$ ($=0.2$ ${\rm Z}_\odot$) 
appropriate for the progenitor of SN 2007bi.
We calculate the evolution of very massive stars with MS mass range
$M_{{\rm MS}} = 100 - 500$ M$_\odot$ taking into account uncertainties in
mass-loss rate.
We show the CO core mass as a function of the MS mass.
Then, we constrain the MS and CO core masses using the amount
of $^{56}$Ni observed in SN 2007bi.
We also discuss the relationship between the SN type and surface He abundance.
Finally, we deduce the ranges of MS mass in PI and CC SN models
appropriate for SN 2007bi and discuss the possible explosion mechanism.

\section{Progenitor Model}

\subsection{Stellar evolution model}

In order to estimate the progenitor mass of SN 2007bi, 
we calculated the evolution of very massive stars from H burning until 
central He exhaustion. 
We considered a mass range of zero-age MS stars of 
$M_{{\rm MS}} = 100 - 500$ M$_\odot$.
The initial metallicity was set to be $Z_0 = 0.004$ ($= 0.2$ Z$_\odot$).
The initial mass fractions of $^1$H and $^4$He were set to be 0.7492 and
0.2468, respectively.
The relative abundances of species heavier than He were derived from 
\citet{Anders89}.

The stellar evolution code was updated from \citet{Saio88} and \cite{Umeda05}.
We used a nuclear reaction network consisting of
282 species of nuclei from n, p, to Br to calculate the time evolution of 
the chemical composition distribution and nuclear energy
generation.
We adopted reaction rates from the current version of 
REACLIB \citep{Rauscher00}.
The reaction rate of $^{12}$C($\alpha,\gamma)^{16}$O was taken 
from \citet{Caughlan88}, multiplied by a factor of 1.6.
For the convection treatment we assumed the Schwarzschild criterion for 
convective instability, and a diffusive treatment for the convective mixing.
The opacity was evaluated using tables of  OPAL opacity
\citep{Iglesias96},  molecular opacity \citep{Ferguson05}, 
and conductive opacity \citep{Cassisi07}.

\subsection{Mass-loss rate}

Since mass-loss rates suffer from various uncertainties, we present
models computed with the following three different mass-loss recipes.

\subsubsection{Case A (standard case)}

The mass-loss rate of MS stars with effective temperature 
$T_{{\rm eff}} >$ 12 000 K and H mass fraction $X \ge 0.4$ 
was taken from \citet{Vink01}.
The mass-loss rate of stars with low surface temperature 
$T_{{\rm eff}} \le$ 12 000 K was adopted from \citet{deJager88}.
We multiplied the mass-loss rate by the metallicity dependence factor 
$(Z/{\rm Z}_\odot)^{0.64}$.
The power index 0.64 is the same value as that of B supergiants in
\citet{Vink01} and is consistent with observational scaling \citep{Mauron11}.

We set the mass-loss rate of Wolf-Rayet (WR) stars with 
$T_{{\rm eff}} >$ 12 000 K and $X < 0.4$ as
\begin{eqnarray}
\log (-\dot{M}) &=& -11.0 + 1.29 \log (L/{\rm L}_\odot) \nonumber \\
&+&1.73 \log {\rm min}(Y_{\rm s},0.98) + 0.47 \log {\rm max}(Z,0.02) 
\nonumber \\
&+&\alpha \log (Z_0/0.02) \quad [\log ({\rm M}_\odot {\rm yr}^{-1})],
\end{eqnarray}
where $L$ and $Y_{\rm s}$ are the luminosity and helium mass fraction at the 
surface.
The factor for the initial metallicity dependence $\alpha$ was set to be
0.86 and 0.66 for WN and WC stars, respectively.
WR stars were classified into WN, WC and WO stars in accordance with
\cite{Georgy09}.
We took into account equation (22) in \citet{Nugis00} and 
the initial-metallicity dependence in \citet{Vink05}.
Extreme, episodic mass loss in WR stars, which has been observed in some SNe
\citep[e.g.][for SN 2006jc]{Foley07} was not taken into account in the
present study.

\subsubsection{Case B (larger mass loss during the WR phase)}

We adopted the same mass-loss rates in the MS and low surface temperature 
phases, and rates with metallicity-dependent indices $\alpha =$ 0.60 
and 0.40 for WN and WC stars, respectively.
The metallicity dependence of the mass-loss rate of WR stars 
in the Large Magellanic Cloud and Small Magellanic Cloud was observed as 
$\propto (Z_0/{\rm Z}_\odot)^{0.8 \pm 0.2}$ 
for WN stars and $\propto (Z_0/{\rm Z}_\odot)^{0.6 \pm 0.2}$ for WC stars
\citep{Crowther07}.
If the scaling index is small, the corresponding mass-loss rate is large.

\subsubsection{Case C (smaller mass loss during whole lifetime)}

We reduced the mass-loss rate in case A by a factor of 2 during the
whole lifetime.
Overestimations of the empirical mass-loss rate in OB stars and WR stars have
been discussed \citep[see the review in][]{Crowther07,Pulse08}.
The reduction of mass-loss rate depends on wind clumping 
\citep[e.g.][]{Fullerton06,Mokiem07} and the degree of clumping has
large uncertainties \citep{Pulse08}.
It is considered that a reduction within a factor of 2 from the theoretical
evaluation \citep{Vink01} is allowed from stellar evolution models
\citep{Hirschi08,Pulse08}.
The effect of wind clumping in WR stars was taken into account by
\citet{Nugis00}.
The reduction of global mass-loss rate of WR stars by wind clumping is
a factor of $2 - 4$ relative to the homogeneous model \citep{Crowther07}.
The surface temperature of very massive stars does not decrease to 
$\log T_{{\rm eff}} < 3.6$, and dust-driven winds might be ineffective.
In this case, wind clumping might be effective for low surface temperature.

\begin{figure}
\begin{center}
\includegraphics[width=8cm]{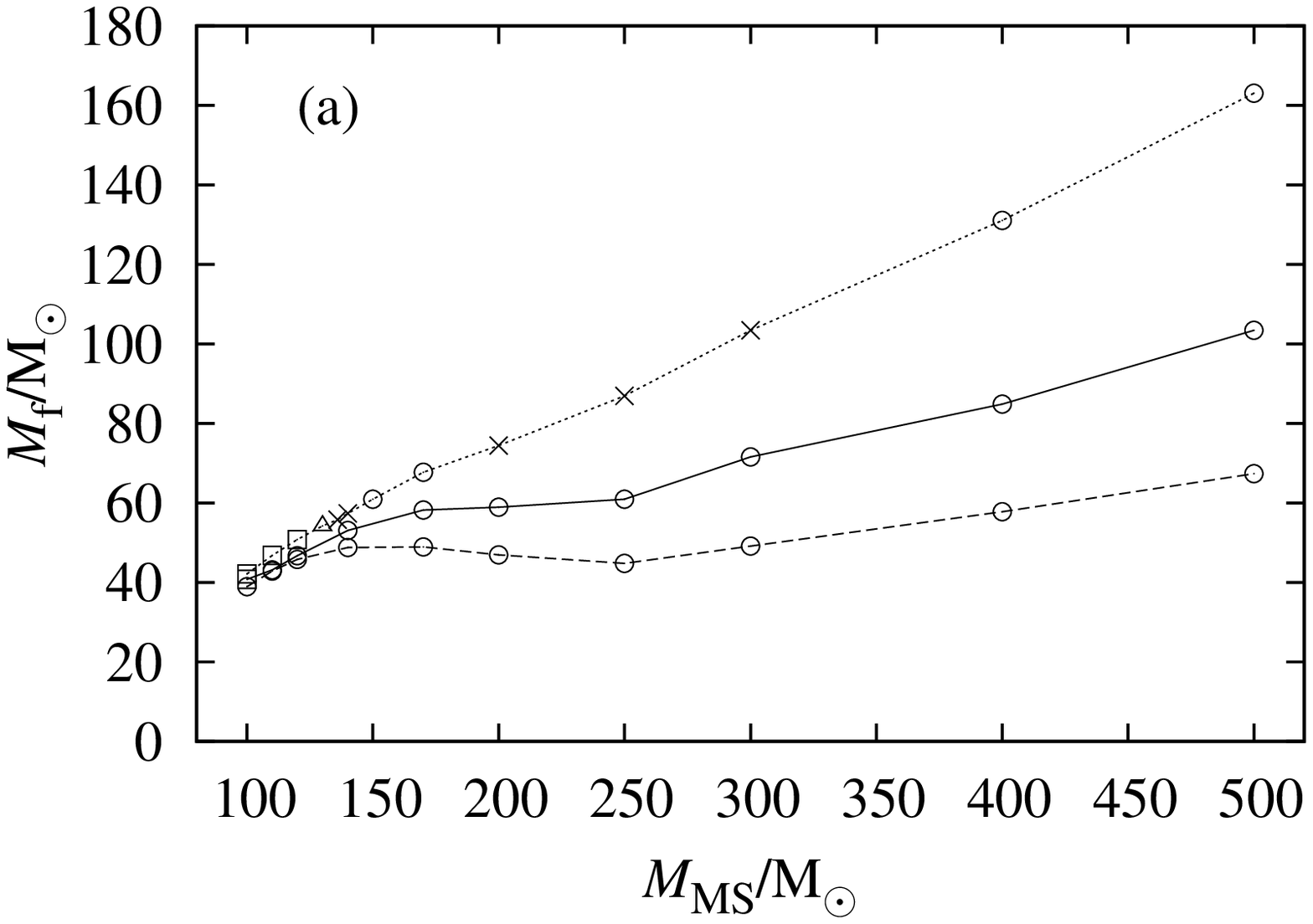}
\includegraphics[width=8cm]{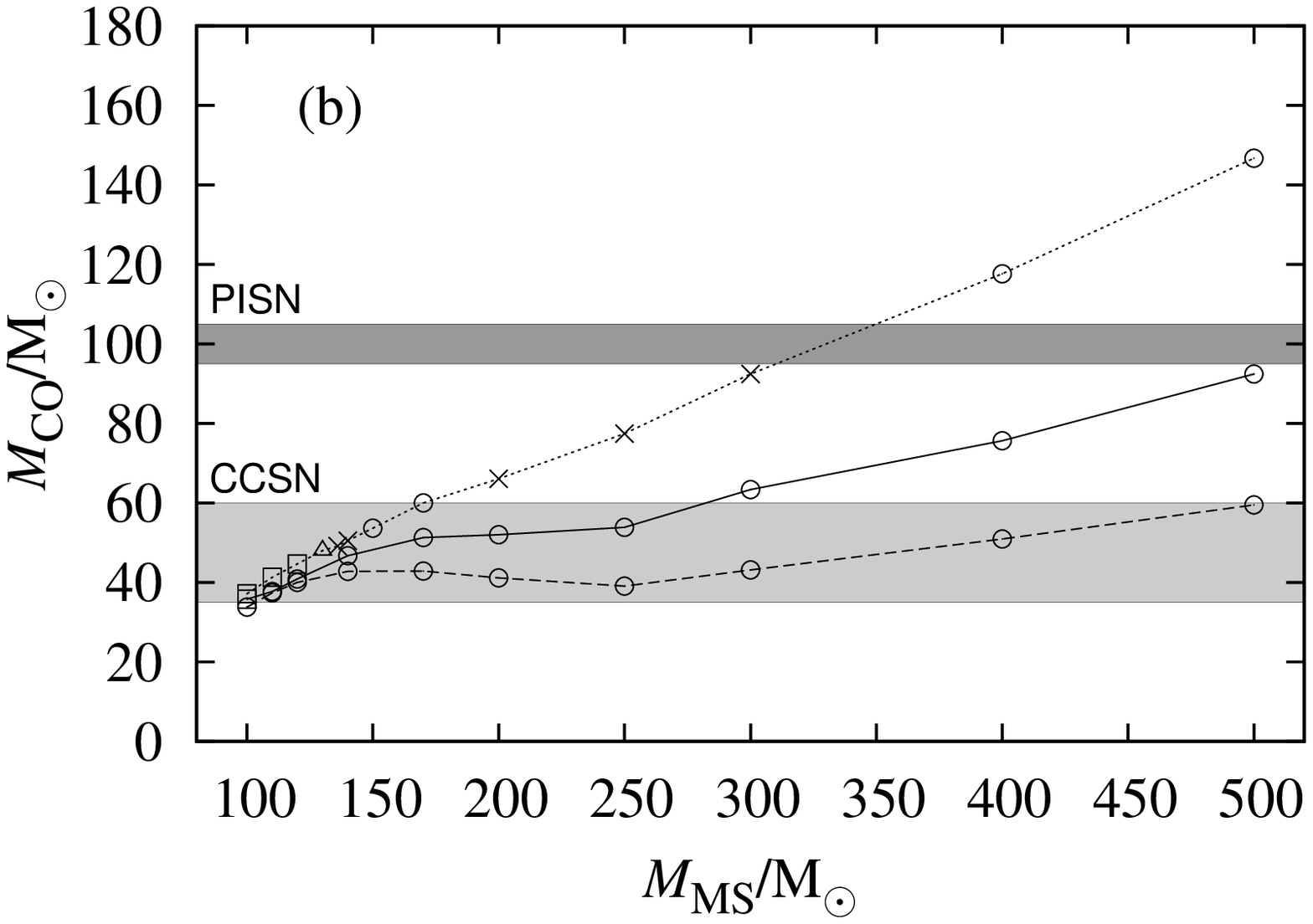}
\end{center}
\caption{
The final mass (a) and CO core mass (b) as a function of the MS mass.
Solid, dashed, and dotted lines correspond to cases A, B and C, respectively.
Squares, triangles, crosses and circles indicate WN, ^^ He-rich' WC 
(see Section 3.3), WC and WO stars.
Dark and light shaded regions denote the mass ranges of the CO cores 
appropriate for SN 2007bi to explode as a PI SN and a CC SN, respectively.
}
\label{fig1}
\end{figure}

\section{Results}

\subsection{Final mass of very massive stars}

Fig. 1(a) shows the final mass $M_{\rm f}$ and type of
WR star as a function of the MS mass.
The final progenitor mass depends on the mass-loss rate as well as on 
the MS mass.
The dependence on the mass-loss rate is larger for more massive stellar
models.
In case A the final mass increases with the MS mass.
These stars evolve to WO stars except for the one with 
$M_{{\rm MS}} = 100$ M$_\odot$.
The final mass of the stars in case B rises with the MS mass
except for the range of 
$170 \le M_{{\rm MS}} \le 250$ M$_\odot$.
All of the stars evolve to WO stars.
The increasing final mass is also shown in case C.
The final type of the star depends on the MS mass.
The relatively ^^ low'-mass stars with 
$100 \le M_{{\rm MS}}$ $\le$ 120 M$_\odot$ evolve to WN stars.
The stars with $130  \le M_{{\rm MS}}$ $\le$ 140 $M_\odot$ and
$200 \le M_{{\rm MS}}$ $\le$ 300 M$_\odot$ evolve to WC stars.
The former WC stars lose their remaining He envelope during carbon burning.
The He envelope of the latter stars has mostly been stripped away before
carbon burning.
The other stars evolve to WO stars.

The mass of the CO core of an evolved star is important for the explosion
mechanism of SN 2007bi.
Fig. 1(b) shows the CO core mass $M_{{\rm CO}}$ as a function of
the MS mass.
The mass of the CO core is defined as the largest mass coordinate satisfying
a He mass fraction smaller than $10^{-3}$ \citep{Umeda08}.
The CO core mass of a star with a given $M_{{\rm MS}}$ and mass-loss rate
is smaller by $\sim 10 - 13$ per cent than the corresponding final mass.

\subsection{Ejected amount of $^{56}$Ni and mass of CO core}

From analyses of the light curve and spectra of SN 2007bi,
the ejection of $3.6-7.4$ M$_\odot$ of $^{56}$Ni has been derived.
The average amount of $^{56}$Ni is 5.3 M$_\odot$.
The amount of ejected $^{56}$Ni has been connected to the CO core mass 
in previous studies.
Here we constrain the range of MS mass from the amount of $^{56}$Ni 
in SN 2007bi.

Nucleosynthesis studies of PI SNe have indicated that PI SN models with 
95 $\la$ $M_{\rm CO}$ $\la$ 105 M$_\odot$ produce 
$3-10$ M$_\odot$ of $^{56}$Ni \citep{Heger02}.
The PI SN models of \citet{Umeda02} obtained a similar result.
Therefore the progenitor of the PI SN corresponding to SN 2007bi 
should be a CO core with 95 $\la M_{\rm CO}$ $\la$ 105 M$_\odot$.

The MS mass range deduced from the CO core mass is shown as the dark shaded 
region in Fig. 1(b).
This criterion is not satisfied in the MS mass range of 
$M_{{\rm MS}} \le 500 {\rm M}_\odot$ in cases A and B.
The MS mass range in case A is estimated to be
515 $\le$ $M_{{\rm MS}}$ $\le$ 575 M$_\odot$ from linear
extrapolation in Fig. 1($b$).
In case C, the appropriate range reduces to
310 $\le$ $M_{{\rm MS}}$ $\le$ 350 M$_\odot$.

The amount of $^{56}$Ni produced in a CC SN model gives a
lower limit on the progenitor mass.
We expect from the result of \citet{Umeda08} that a SN with a
kinetic energy of $\sim 3 \times 10^{52}$ erg can produce more than 
3 M$_\odot$ of $^{56}$Ni if the CO core of the progenitor is larger than
$\sim 35$ M$_\odot$.
The upper limit of the MS mass might be the lowest mass of a PI SN progenitor.
Theoretical studies of PI SNe have indicated that a CO core larger than 
$\ga 60$ M$_\odot$ explodes as a PI SN \citep{Heger02,Umeda02}.
We consider the upper limit of a CC SN progenitor as 
$\sim 60$ M$_\odot$.
Therefore, the CO core mass appropriate for explaining SN 2007bi with a CC
explosion is 35 $\la$ $M_{{\rm CO}}$ $\la$ 60 M$_\odot$.

The range of MS mass for CC SN models is shown as the light shaded 
region
in Fig. 1(b).
The range extends from 100 to 280 M$_\odot$ in case A.
In case B, all models except for $M_{{\rm MS}} = 100$ M$_\odot$ will 
explode as CC SNe appropriate for SN 2007bi.
On the other hand, in case C, the mass range is
limited to 100 $\la$ $M_{{\rm MS}}$ $\la$ 170 M$_\odot$.

\subsection{Surface He abundance for SNe Ib/Ic}

SNe Ic are characterized by weak or absent He spectra.
However, a quantitative criterion to distinguish between SNe Ib and Ic has not
been theoretically established.
The He lines are considered to appear because of the excitation
of He by non-thermal electrons excited by $\gamma$-rays from the decays of
$^{56}$Ni and $^{56}$Co \citep{Lucy91}.
The strength of the He lines should be sensitive to the amounts of He and 
$^{56}$Ni, the amount of matter in intermediate layers between Ni and He 
layers which attenuate the $\gamma$-rays, the degree of mixing of Ni into 
He layer, etc.
Criteria to distinguish between SNe Ib and Ic were discussed 
using the total He mass of a progenitor \citep{Wellstein99,Georgy09,Yoon10} or
the He mass fraction at the outermost layers \citep{Yoon10}.
The effects of thickness of the intermediate layers
and the degree of mixing have been investigated \citep{Woosley97}.
We discuss the possibility of a SN Ic progenitor by considering the 
total He mass,
the He mass fraction at the surface and the mass ratio of He to the
intermediate layers.

The first criterion is based on the total He mass.
We consider two cases for the He mass limit of SNe Ic:
0.5 and 1.5 ${\rm M}_\odot$.
In previous studies, the He mass limit of SNe Ic was assumed to be 
0.5 ${\rm M}_\odot$ \citep{Wellstein99,Yoon10} or 0.6 ${\rm M}_\odot$
\citep{Georgy09}.
On the other hand, \citet{Georgy09} reported that the choice
of the total He mass
limit between 0.6 and 1.5 M$_\odot$ hardly affects 
the $M_{\rm MS}$ ranges for SNe Ib/Ic.
\citet{Yoon10} suggested in discussion that
He lines are not seen in early-time spectra even though the total He mass is
as large as 1.0 M$_\odot$ if He is well mixed with CO material 
having $Y_{\rm s} \la 0.5$.
Fig. 2(a) shows the total He mass versus the MS mass.
When the He mass limit is 0.5 ${\rm M}_\odot$, the MS mass is limited
in the very narrow range of $110-120$ M$_\odot$ in case A and
$100-115$ M$_\odot$ in case B.
In case C no progenitors explode as SNe Ic.
When the He mass limit is 1.5 ${\rm M}_\odot$, all progenitors except for
WN stars, an ^^ He-rich' WC star with $M_{{\rm MS}} = 130$ M$_\odot$
in case C, and stars with $M_{{\rm MS}} > 350$ M$_\odot$ in case C
will explode as SNe Ic.

\begin{figure}
\includegraphics[width=80mm]{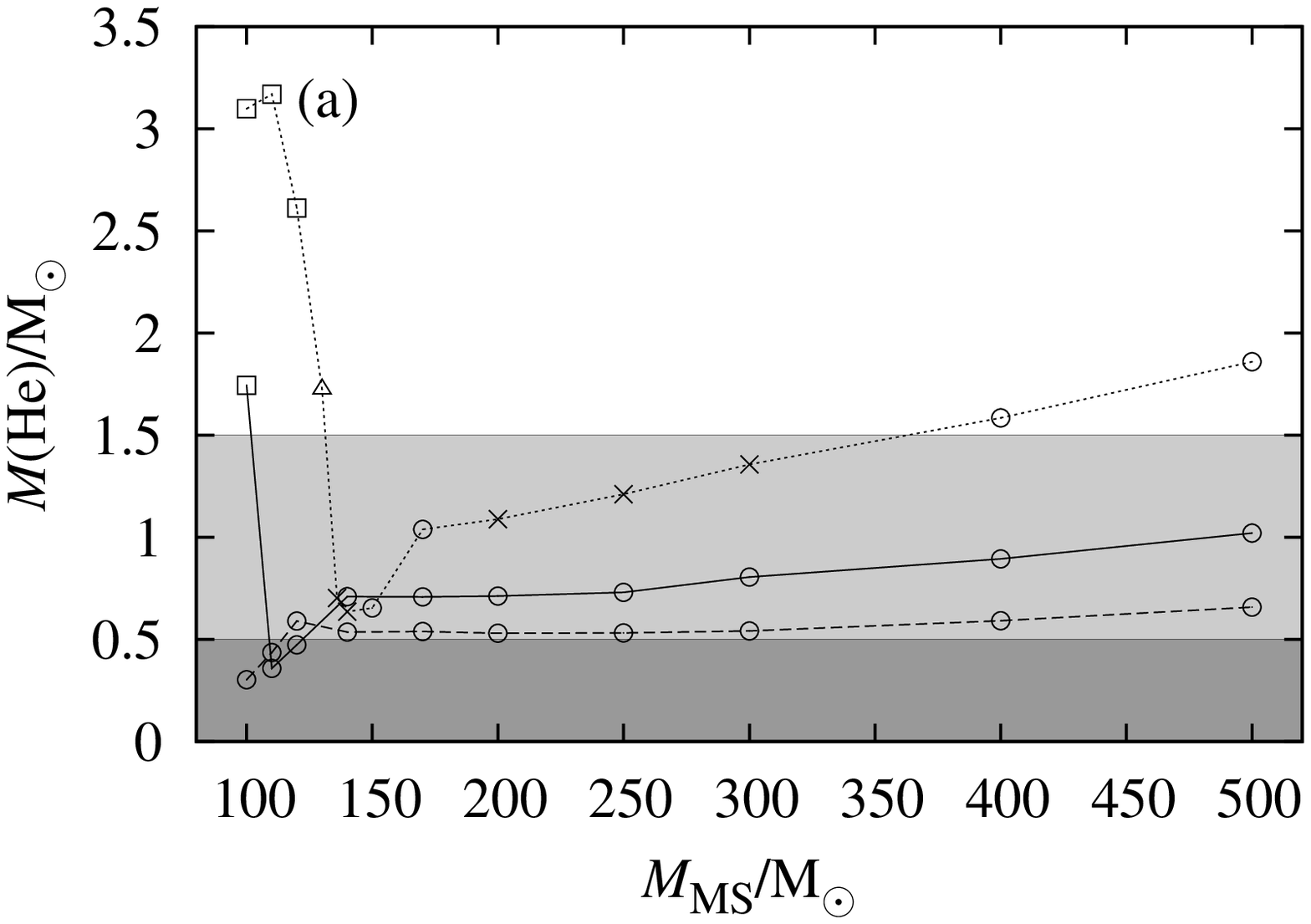}
\includegraphics[width=80mm]{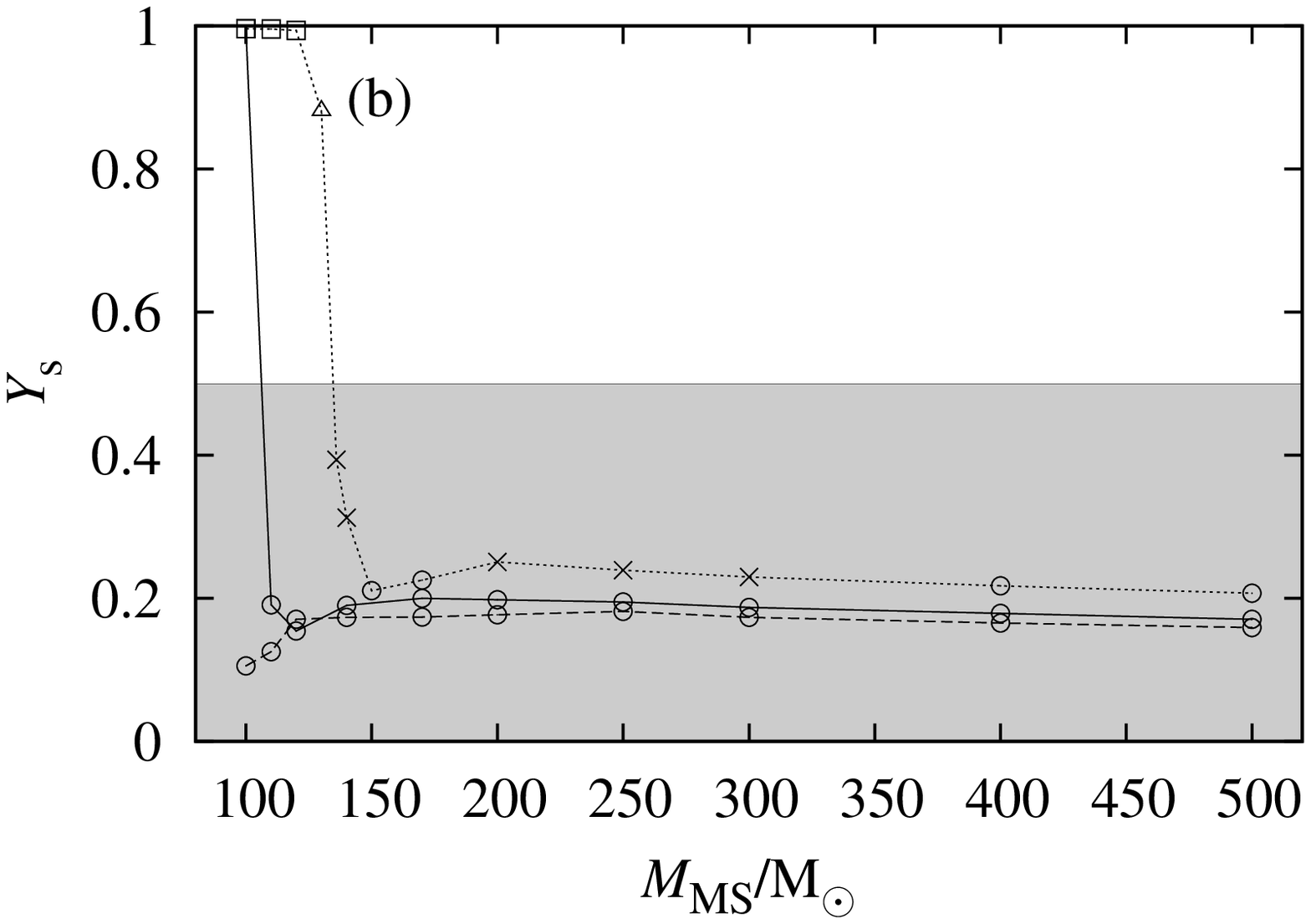}
\includegraphics[width=80mm]{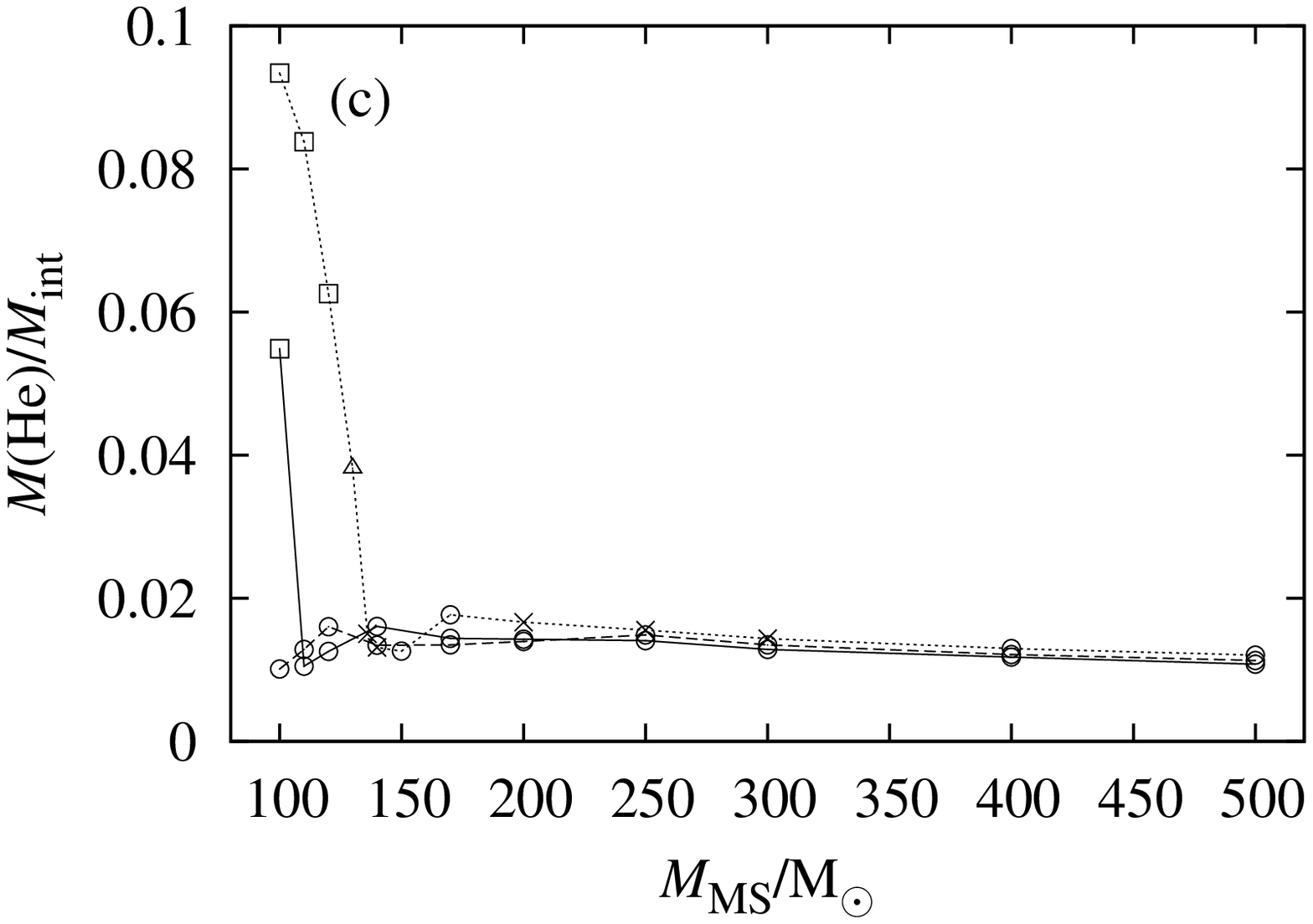}
\caption{The total He mass in the envelope (a), the He mass fraction 
at the surface (b), and the mass ratio of He to the intermediate layers
(c) plotted against the MS mass.
Solid, dashed and dotted lines correspond to cases A, B and C, respectively.
Squares, triangles, crosses and circles indicate WN, ^^ He-rich' WC, WC 
and WO stars.
In panel (a), dark and light shaded regions satisfy the criteria  
$M({\rm He}) \le 0.5$ and 1.5 M$_\odot$, respectively.
In panel (b), the shaded region satisfies the criterion $Y_{\rm s} \le 0.5$.
}
\label{fig2}
\end{figure}

The second criterion is the He mass fraction at the surface.
We set this criterion as $Y_{\rm s} \le 0.5$ in accordance with the discussion
in \citet{Yoon10} as mentioned above.
Fig. 2(b) presents the He mass fraction at the surface as a function of
the MS mass.
All WO stars and WC stars except for the He-rich WC will explode as SNe Ic.

We also consider the mass ratio of He to the intermediate layers.
The mass of the intermediate layers is assumed to be 
$M_{{\rm int}} = M_{\rm f} - 9.0$ M$_\odot$.
The mass of 9.0 ${\rm M}_\odot$ corresponds to the amount of ejected
$^{56}$Ni and the central remnant in the CC SN model.
The masses of $^{56}$Ni and the central remnant were
evaluated to be 6.1 and 3.0 ${\rm M}_\odot$, respectively, 
for the $43-$M$_\odot$ progenitor \citep{Moriya10}.
Although the amount of $^{56}$Ni would be smaller than 9 ${\rm M}_\odot$ 
and there is no remnant in the PI SN model, the difference is not important 
because the final mass of the PI SN progenitors is $\sim 100$ M$_\odot$.
The mass ratio of He to the intermediate layers is shown in Fig. 2(c).
The ratios of WN stars and the He-rich WC star are clearly larger than
those of WO and the other WC stars.
We may set the criterion of the ratio as $\sim 0.02 - 0.03$.
In this case, the range of MS mass is the same as that deduced from
the criterion of the He mass fraction at the surface.

\begin{table}
\caption{
The range of MS mass for PI and CC SN models appropriate for SN 2007bi
deduced from the mass range of the CO core and the surface He abundance.
The ratio of the probability of explosion as a PI SN to explosion as a CC SN 
for SN 2007bi, $r_{{\rm PI/CC}}$, is also listed (see Section 4).
}
\begin{scriptsize}
\begin{center}
\begin{tabular}{lccc}
\hline
Condition & PI SN & CC SN & $r_{{\rm PI/CC}}$ \\
 & (M$_\odot$) & (M$_\odot$) & \\
\hline
\multicolumn{4}{c}{Case A} \\
$M({\rm He}) \le 0.5 {\rm M}_\odot$ & $-$ & $110 - 120$ & 0 \\
$M({\rm He}) \le 1.5 {\rm M}_\odot$ or $Y_{\rm s} \le 0.5$ &
$515 - 575$ & $110 - 280$ & 0.024 \\
\multicolumn{4}{c}{Case B} \\
$M({\rm He}) \le 0.5 {\rm M}_\odot$ & $-$ & $110 - 115$ & 0 \\
$M({\rm He}) \le 1.5 {\rm M}_\odot$ or $Y_{\rm s} \le 0.5$ &
$-$ & $110 - 500$ & 0 \\
\multicolumn{4}{c}{Case C} \\
$M({\rm He}) \le 0.5 {\rm M}_\odot$ & $-$ & $-$ & $-$ \\
$M({\rm He}) \le 1.5 {\rm M}_\odot$ or $Y_{\rm s} \le 0.5$ &
$310 - 350$ & $135 - 170$ & 0.19 \\
\hline
\end{tabular}
\end{center}
\end{scriptsize}
\end{table}

\section{Range of Main-Sequence Mass for the SN 2007bi Progenitor}

We evaluate the ranges of MS mass appropriate for explaining SN 2007bi
by the PI SN and CC SN models taking into account the range of CO core mass 
and the surface He abundance.
Table 1 presents the MS mass range with different conditions of
He abundance and mass-loss rate.
The MS mass range strongly depends on the surface He conditions.
If He lines appear with a total He mass larger than 0.5 M$_\odot$, then there
is no mass range of the PI SN model for SN 2007bi.
The MS mass range of the CC SN model is also limited to 
$\sim 110 - 120$ M$_\odot$.
However, this condition may be too strict, as discussed by \citet{Yoon10}.
If the progenitors explode as Type Ic SNe with 
$M({\rm He)} \le 1.5$ M$_\odot$ or $Y_{\rm s} \le 0.5$, the MS mass range 
is mainly determined by the condition of the CO core mass.
Thus the minimum MS mass for the PI SN model is 310 M$_\odot$.
The MS mass range of the CC SN model in case A becomes slightly narrower:
110 $\le$ $M_{{\rm MS}}$ $\le$ 280 M$_\odot$.
In cases B and C the mass range is determined by the mass of the CO core.

We now discuss the probabilities of SN 2007bi exploding as a PI SN and 
a CC SN, considering the MS mass range and Salpeter initial 
mass function (IMF) ($dN/dM_{{\rm MS}} \propto M_{{\rm MS}}^{-2.35}$).
The ratio of the probability of explosion as a PI SN to explosion as a CC SN, 
$r_{{\rm PI/CC}}$, is also listed in Table 1.
If He lines are not seen for $M({\rm He}) \le 1.5$ M$_\odot$ or 
$Y_{\rm s} \le 0.5$,
the probability of the CC SN model is about 40 times larger than that of 
the PI SN model in case A.
In case B, the PI SN model is not adequate for SN 2007bi because of large 
mass loss.
In case C, on the other hand, the probability of the CC SN model is larger 
than that of the PI SN model by a factor of 5.
However, if He lines are observed with a total He mass of 
$M({\rm He}) > 0.5$ M$_\odot$, the PI SN model is inadequate for 
SN 2007bi in all mass-loss cases.
It is important to establish more definite classification criteria for SNe Ic.

It has been argued that very massive CO cores would 
collapse to form black holes and become dark or very faint SNe 
\citep[e.g., review in][]{Smartt09}.
In this case, the probability of explosion as a CC SN would become much 
smaller.
Observational or theoretical evaluation of the fraction of dark
or faint SNe in the death of very massive stars will more precisely
constrain the probability of the CC SN model.

\section{Concluding Remarks}

We have investigated the evolution of very massive stars 
with $M_{{\rm MS}} = 100-500$ M$_\odot$ and $Z_0 = 0.004$ from 
H burning to central He exhaustion to constrain the progenitor of 
an extremely luminous Type Ic SN 2007bi.
If SN 2007bi is a PI SN, the progenitor could be evolved from a MS star with 
515 $\la$ $M_{{\rm MS}}$ $\la$ 575 M$_\odot$.
If SN 2007bi is a CC SN, the appropriate range of the MS mass is
110 $\la$ $M_{{\rm MS}}$ $\la$ 280 M$_\odot$.
When we take into account a Salpeter IMF, the probability of SN 2007bi 
exploding as a CC SN is about 40 times larger than for a PI SN.
If the mass-loss rate is small,
the minimum mass for the progenitor becoming a PI SN is about 310 M$_\odot$.
The mass range changes to $M_{{\rm MS}} \sim 310-350$ M$_\odot$
for the PI SN model and $M_{{\rm MS}} \sim 135-170$ M$_\odot$ for 
the CC SN model.
If the mass-loss rate is large or He lines appear for an He mass
of $M({\rm He}) > 0.5$ M$_\odot$, then SN 2007bi should have exploded 
as a CC SN.
Although the light curve of SN 2007bi was fitted by both the
PI SN and CC SN models, the CC SN model is favoured as explaining 
the explosion of SN 2007bi from arguments based on the probability ratio 
of the PI SN model appropriate for SN 2007bi to the CC SN model.

We should note that the conditions of He abundance for the progenitor 
to become a Type Ic SN strongly affect the probability of a PI SN.
If a total He mass $M({\rm He}) > 0.5$ M$_\odot$ enables the formation
of He spectra, the PI SN model would be inadequate for SN 2007bi even 
for a small mass-loss rate.
It is quite important to evaluate more definite criteria to classify
SNe Ic.
We also note the probability of direct collapse without bright SNe 
for CC SN cases.
This would reduce the probability of a CC SN for SN 2007bi.
It is necessary to evaluate theoretically 
or observationally the fraction of black hole formation without 
bright SNe in CC SNe.

If the progenitor of SN 2007bi or a more massive star was rotating very fast,
it could explode as a SN Ic associated with a GRB.
We may not have seen such a GRB-SN because of the orientation effect.
This is also interesting point to investigate in the future.

\section*{Acknowledgments}

We acknowledge the anonymous referee for many valuable comments.
We thank Hideyuki Saio for providing the stellar evolution code
and for useful comments.
We are grateful to Masaomi Tanaka, Nobuyuki Iwamoto, and Ken'ichi Nomoto 
for useful discussions.
This work was supported by the grants-in-aid for Scientific Research
(20041005, 20105004) from the MEXT of Japan.

%


\label{lastpage}

\end{document}